\documentclass[letterpaper,10pt]{article} 

\usepackage{opticameet3} 

\newcommand\authormark[1]{\textsuperscript{#1}}

\usepackage{amsmath,amssymb}
\usepackage{subcaption}
\usepackage{mathrsfs}
\usepackage{graphicx}
\usepackage[colorlinks=true,bookmarks=false,citecolor=blue,urlcolor=blue]{hyperref} 
\usepackage[sorting=none]{biblatex}
\begin{document}

\title{Noise-resilient approach for deep tomographic imaging}


\author{Zhen Guo\authormark{1},
Zhiguang Liu\authormark{2}, Qihang Zhang\authormark{1}, George Barbastathis\authormark{2,3,7}, Michael E. Glinsky\authormark{4}} 

\address{\authormark{1}Department of Electrical Engineering and Computer Science,
Massachusetts Institute of Technology,
Cambridge, Massachusetts, 02139,
USA\\
\authormark{2}Department of Mechanical Engineering,
Massachusetts Institute of Technology,
Cambridge, Massachusetts, 02139,
USA\\
\authormark{3}Singapore-MIT Alliance for Research and Technology (SMART) Centre, Singapore 138602\\
\authormark{4}qiTech, Inc., Santa Fe, New Mexico 87505, USA\\}

\email{\authormark{7} gbarb@mit.edu} 

\begin{abstract}
We propose a noise-resilient deep reconstruction algorithm for X-ray tomography. Our approach shows strong noise resilience without obtaining noisy training examples. The advantages of our framework may further enable low-photon tomographic imaging.
\end{abstract}

\section*{Main Text}
X-ray tomography is a non-destructive imaging technique that visualizes the interior features of solid objects, with applications in many disciplines. A high quality reconstruction can be generated by densely-sampled, full-angle, and noise-free measurements covering the Fourier space~\cite{stark1981direct}. However, acquiring such measurements can be a challenge in practice for an object with geometric constraints or radiation sensitivity. For measurements that are limited-angle, sparse-view and noisy, the resulting inverse problem becomes ill-conditioned due to the deficits in the Fourier-space information~\cite{rantala2006wavelet, wurfl2016deep}. Under such sampling conditions, it is essential to utilize a regularization prior on the object distribution to retrieve a high-fidelity reconstruction~\cite{engl1996regularization}.

Conventionally, the regularization prior is a general-purpose penalty function included in the objective function of an iterative optimization algorithm. The final reconstruction is regarded as the maximum a posteriori (MAP) estimate given the prior distribution. Alternatively, regularization priors can also be learned through supervised machine learning method~\cite{jin2017deep, wang2020deep, guo2022randomized}. Given a training dataset, a machine learning algorithm can be trained to create an inverse map that directly generates object reconstructions from the measurements, learning the spatial correlations of the objects implicitly. Some efforts also utilize a conventional algorithm that generates an approximate estimate of the object for a deep neural network, achieving better reconstruction quality by separating the forward model of tomography from the network training~\cite{kang2021dynamical, guo2022physics}.


One major assumption in previous studies of the learned prior is that the noise statistics in the measurements are comparable between training and testing dataset. Assuming measurements are corrupted by Poisson noise, the reconstruction quality from the learned prior is generally accessed by testing data with the same photons per ray as the training data. However, in practical tomography systems, training and testing data might have different number of photons per ray due to the variability in the light source and detector. When the noise statistics in the dataset shifts between training and testing, the generalization of learning is not guaranteed, leading to performance degradation or failure in producing high-fidelity reconstruction~\cite{antun2020instabilities, wu2020stabilizing}. The ability of maintaining reconstruction quality despite variations of the noise statistics in the test data is called the ``noise resilience'' of the learned prior.

\begin{figure}[htbp!]
    \centering
    {{\includegraphics[width=7.1cm]{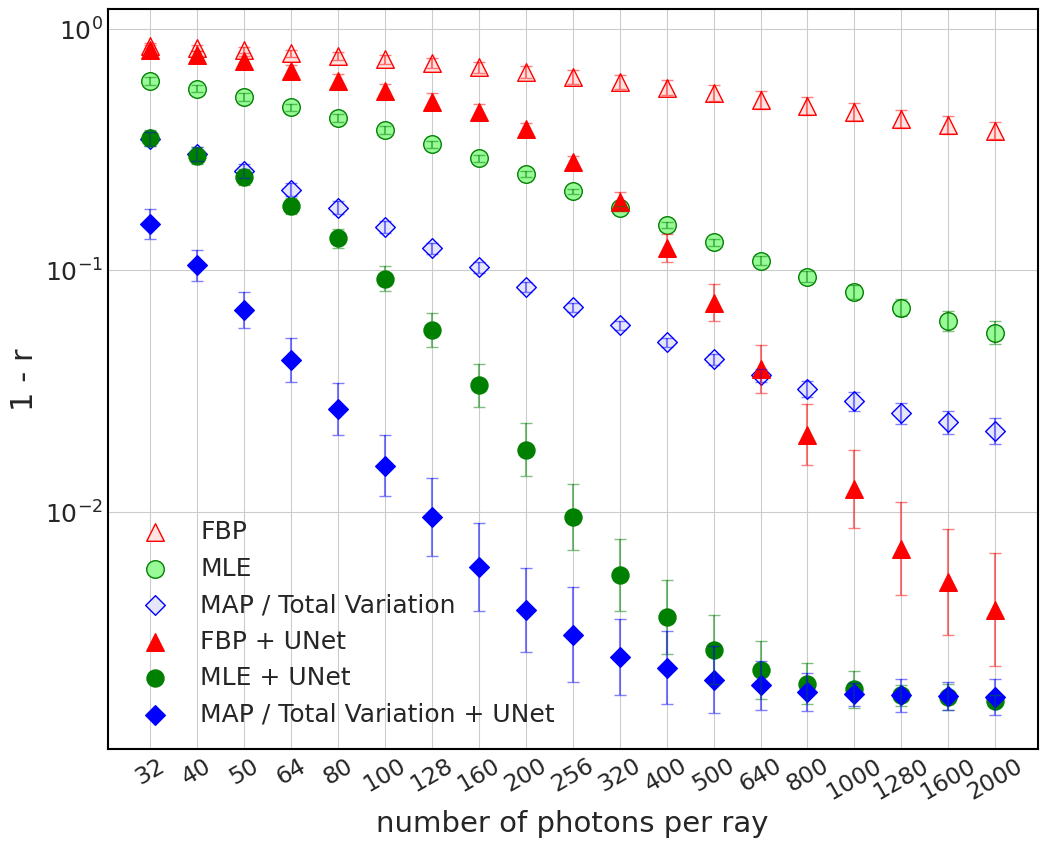} }}%
    \qquad
    {{\includegraphics[width=7.1cm]{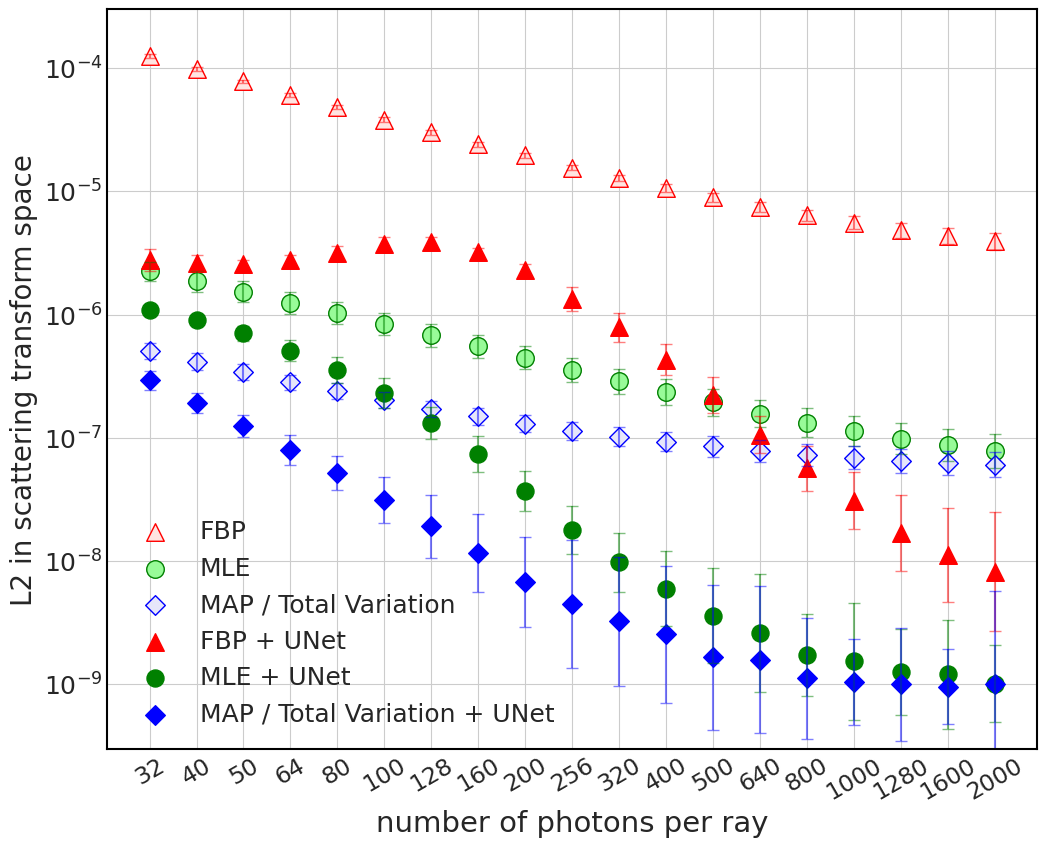} }}%
    \caption{Quantitative comparison between different reconstruction algorithms for tomographic simulations under different photon numbers per ray.  The $x$ axis is the number of photons per ray, and the $y$ axis on the top figure is $1-r$ where $r$ is the Pearson correlation coefficient. The $y$ axis on the bottom is the $L^2$ distance in the scattering transform space.  Here the error-bars are the standard deviation in the log scale.}%
    \label{fig:all-algo1}%
\end{figure}

In this paper, we propose a noise-resilient
deep-reconstruction algorithm for X-ray tomography. Our method improves the noise resilience of the learned prior by more computation, rather than more training data collected under various noise levels. Instead of approximate reconstructions, the inputs to the neural network are Maximum A Posteriori (MAP) Estimate reconstructions from an iterative algorithm. For the special case when the prior distribution is uniform, MAP can be regarded as Maximum Likelihood Estimate (MLE). By enforcing data-fidelity and sparsity prior, the distribution shift between input reconstructions from noise-free and noisy measurements is reduced. Therefore, the noise resilience of the learned prior improves without acquiring the noisy training data.

The quantitative comparison of different algorithms using simulation data is shown in Fig.~\ref{fig:all-algo1}. The imaging condition is constrained to full-angle, sparse sampling (32 projections) and low-photon tomographic simulations where the ill-conditioning becomes severe without regularization. There are three traditional reconstruction algorithms namely FBP, MLE, MAP with TV (total variation), and three learning-based algorithms obtained by adding a UNet for each, designated as
FBP+UNet, MLE+UNet, and MAP+UNet. All the learning-based algorithms share the same optimization parameters and UNet architecture. The Pearson correlation coefficient is to evaluate the perceptual quality of the reconstruction. The scattering transform is to evaluate the morphology of the reconstruction~\cite{mallat2012group}. We report the means and standard errors over the 1000 testing data in the plots. Assuming 0.9 as the acceptable quality for $r$, MAP+UNet satisfies the requirement for the entire photon range, except for 32 photons per pixel.  MLE+UNet is the second best algorithm, which can satisfy the requirement down to 100 photons per pixel. FBP+UNet shows the least noise-resilient among the learned algorithm, withstanding noisy measurements down to 500 photons per pixel. The scattering transform metrics shows similar trend as the Pearson correlation metric, except that the distances between MLE and MAP with TV are relatively small.

\printbibliography

\end{document}